\documentclass{appolb}
\usepackage{graphicx}
\usepackage{epstopdf}

\newcommand{\be}{\begin{equation}}
\newcommand{\ee}{\end{equation}}
\newcommand{\ba}{\begin{eqnarray}}
\newcommand{\ea}{\end{eqnarray}}
\newcommand{\eq}{Eq.~}

\newcommand{\fig}{Fig.~}

\def\lsi{\raise0.3ex\hbox{$<$\kern-0.75em\raise-1.1ex\hbox{$\sim$}}}
\def\gsi{\raise0.3ex\hbox{$>$\kern-0.75em\raise-1.1ex\hbox{$\sim$}}}
\newcommand{\lsim}{\mathop{\lsi}}
\newcommand{\gsim}{\mathop{\gsi}}

\begin{document}
\title{Status of the QCD phase diagram from lattice calculations%
\thanks{Presented at ``Three Days on Quarkyonic Island", HIC for FAIR workshop and 
XXVIII Max Born Symposium, 19-21 May 2011.}%
}
\author{Owe Philipsen
\address{Institut f\"ur Theoretische Physik, Goethe-Universit\"at, Max-von-Laue-Str.~1, \\
60438 Frankfurt am Main, Germany}
}
\maketitle
\begin{abstract}
The present knowledge of the QCD phase diagram based on simulations of lattice QCD
is summarised. The main questions are whether there is a critical point in the QCD
phase diagram and whether it is related to a chiral phase transition. It is shown that
QCD at imaginary chemical potentials has a rich phase structure, which can be determined
in a controlled way without sign problem and which severely constrains the phase structure
at real chemical potentials. 
\end{abstract}
  
\section{Introduction}

\begin{figure}[t]
\vspace*{-0.5cm}

\hspace*{0.3cm}
\includegraphics[width=0.45\textwidth]{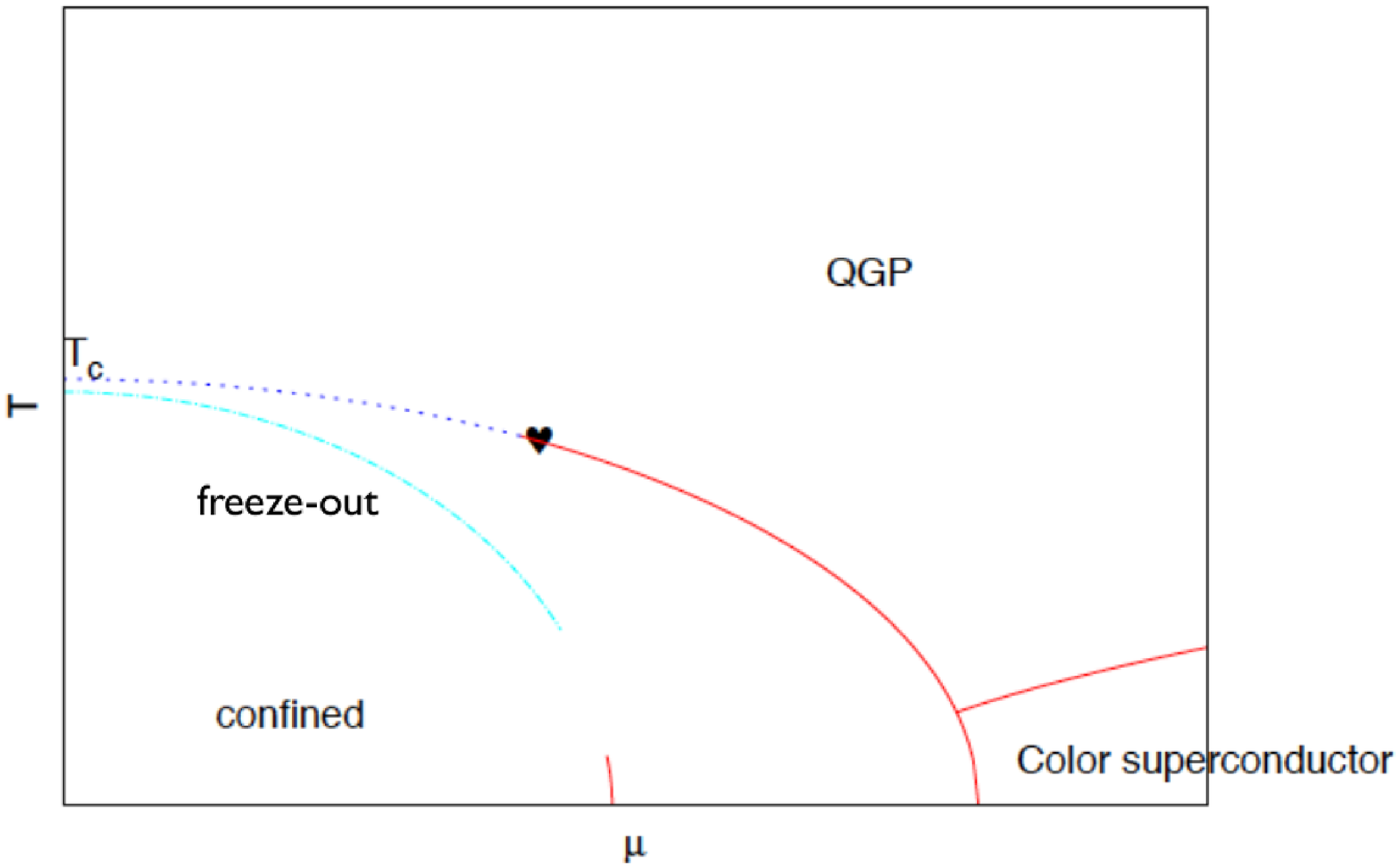}
\includegraphics[width=0.45\textwidth]{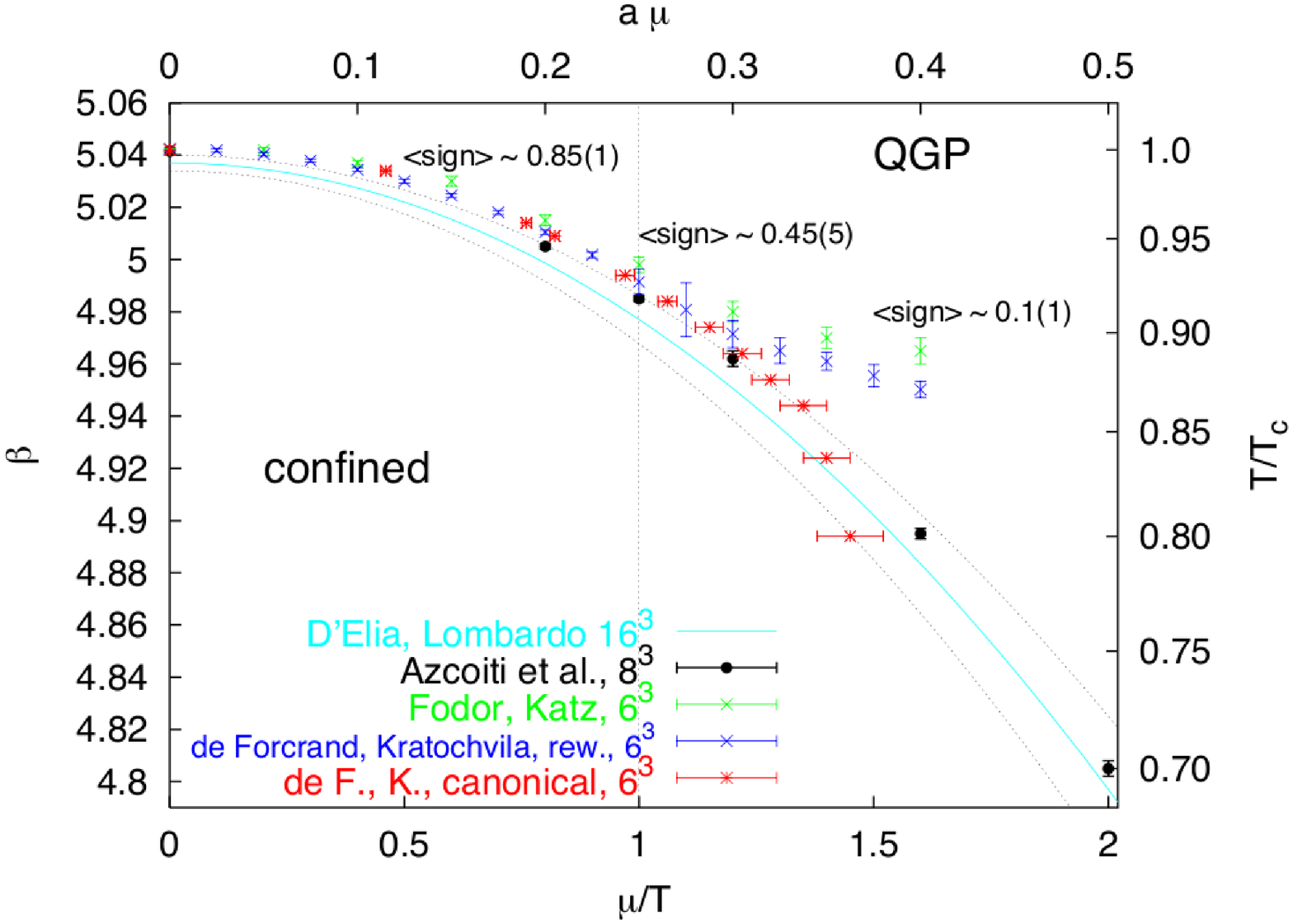}
\caption[]{\label{tc} Left: Sketch of the QCD crossover line $T_c(\mu)$ vs.~the experimental freeze-out curve, which has a larger curvature.  The remaining transition structure is suggestive and not yet conclusive in QCD.
Right: Comparison of the phase boundary for QCD with $N_f=4$ staggered quarks on 
$N_t=4$ lattices. From \cite{slavo}.  }
\end{figure}

The QCD phase diagram determines the form of nuclear matter under different conditions 
as a function of temperature, $T$, and chemical potential for baryon number, $\mu_B$.
Based on asymptotic freedom, one expects at least 
three different forms of nuclear matter: hadronic (low $\mu_B,T$), quark gluon plasma (high $T$) 
and colour-superconducting (high $\mu_B$, low $T$), as sketched in \fig\ref{tc} (left). 
Whether and where these regions are separated
by true phase transitions has to be determined by first principle calculations and experiments.
Since QCD is strongly coupled on scales of nuclear matter, a non-perturbative treatment is 
necessary and Monte Carlo simulations
of lattice QCD are a natural approach. Unfortunately, the so-called sign problem prohibits straightforward
simulations at finite bary\-on density. There are several approximate ways to circumvent this problem, 
all of them valid for $\mu/T\lsim1$ only \cite{oprev,csrev} (with quark chemical potential $\mu=\mu_B/3$): reweighting,
Taylor expansions in $\mu/T$ about zero and simulations at imaginary chemical potential,
where there is no sign problem. The latter can be either analytically continued or used as input
for the canonical partition function.
As long as $\mu/T\lsim1$, all give
quantitatively agreeing results for observables. A direct comparison for the calculation of  the phase
boundary $T_c(\mu)$ for a theory with four flavours is shown in \fig\ref{tc} (right) \cite{slavo}. Note that,
on finite volumes, a transition is always analytic and the phase boundary only pseudo-critical.
Determining the order of the transition in the thermodynamic limit, 
and hence the existence of a chiral critical point, requires costly finite size scaling analyses  
and is a much harder task.

\begin{figure}[t]
\includegraphics[width=0.3\textwidth]{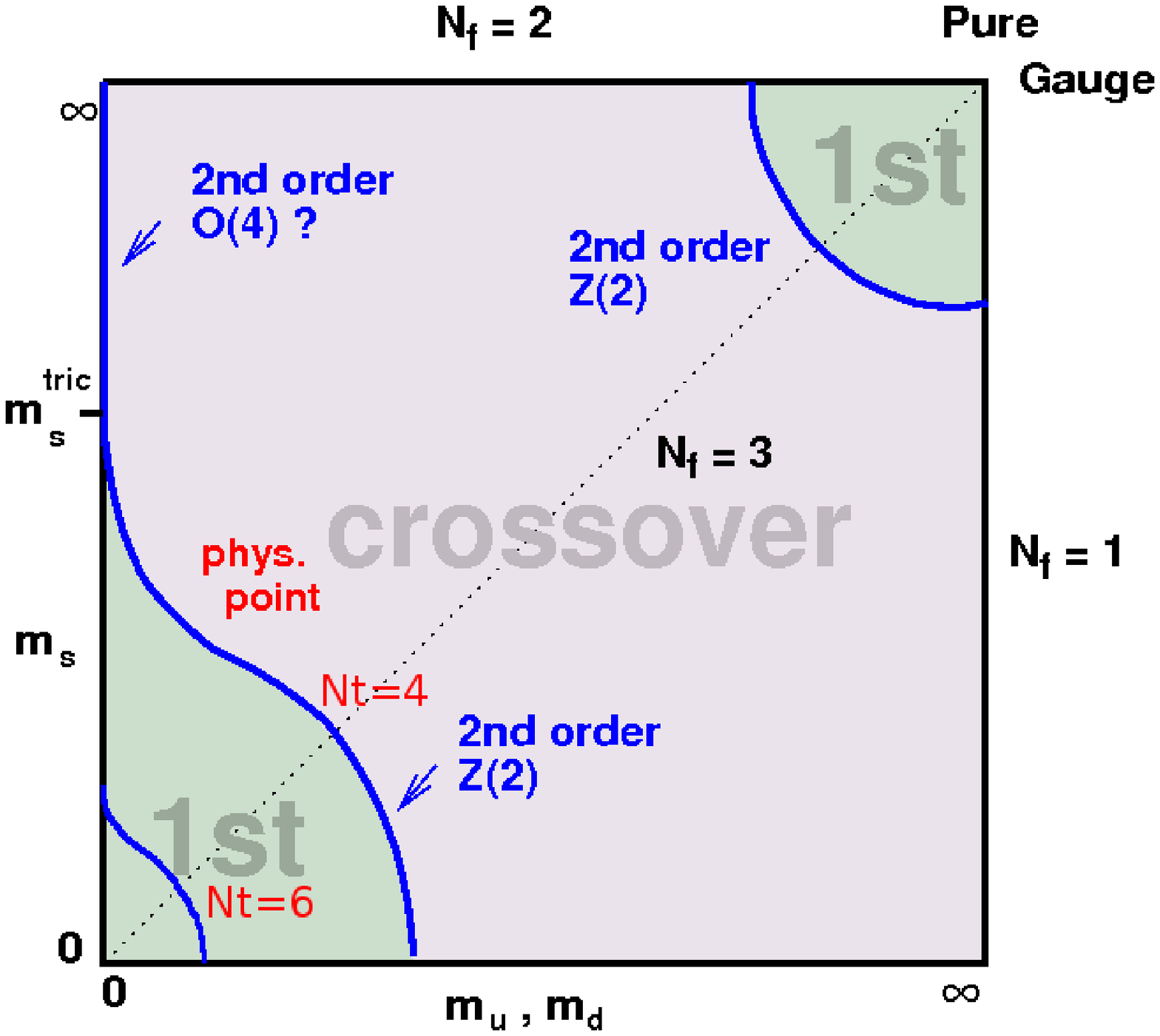}
\includegraphics[width=0.34\textwidth]{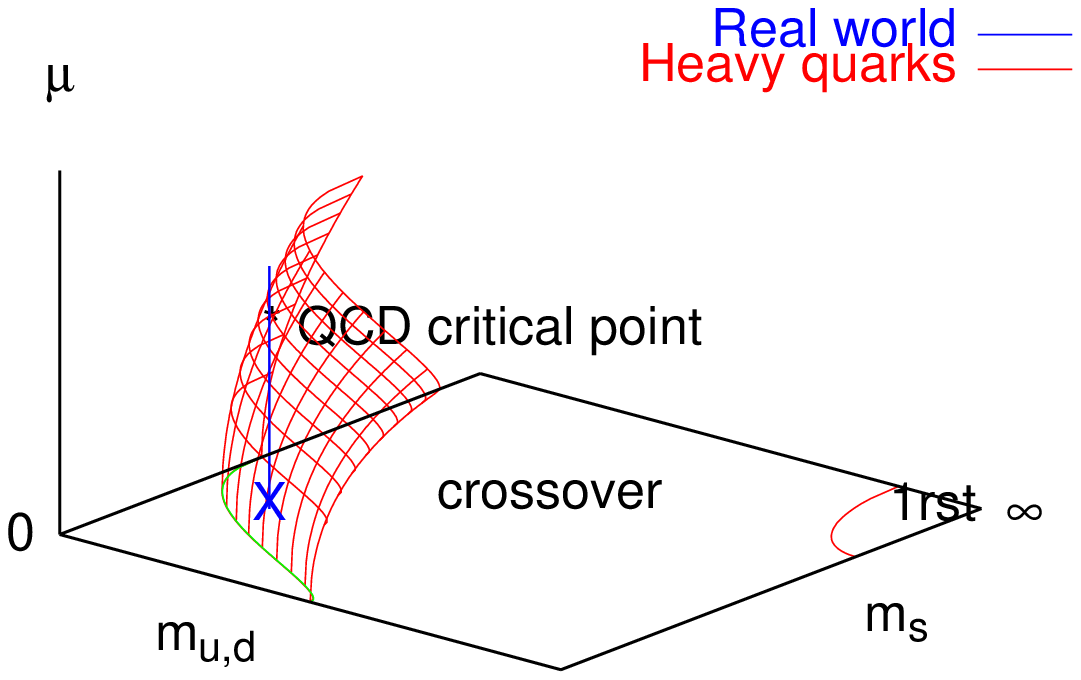}
\includegraphics[width=0.34\textwidth]{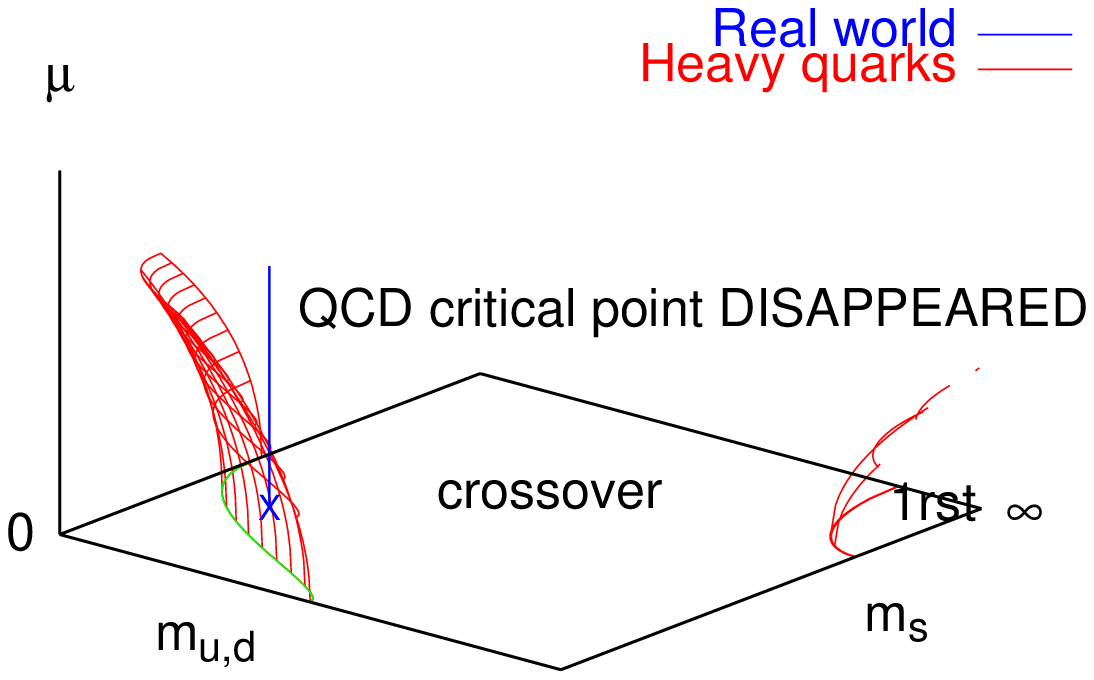}
\caption{\label{schem} Left: Schematic phase transition behaviour of $N_f=2+1$
QCD for different choices of quark masses at
$\mu=0$. On finer lattices, the chiral critical line moves towards smaller quark masses. 
Middle + Right: Chiral critical surface swept by the chiral 
critical line as $\mu$ is turned on. Depending on its curvature, a QCD chiral critical
point is present or absent. 
}
\end{figure}

The order of the finite temperature phase transition at zero density depends on the quark masses and
is schematically shown in \fig\ref{schem} (left).
In the limits of zero and infinite quark masses (lower left and upper 
right corners), order parameters corresponding to the breaking of a 
global symmetry can be defined, and for three degenerate quarks 
one numerically finds first order phase
transitions at small and large quark masses at some finite
temperatures $T_c(m)$. On the other hand, one observes an analytic crossover at
intermediate quark masses, with second order boundary lines separating these
regions. Both lines have been shown to belong to the $Z(2)$ universality class
of the 3d Ising model \cite{kls,fp2,kim1}. 
A convenient observable is the Binder cumulant for the order parameter
$B_4(X) \equiv \langle (X - \langle X \rangle)^4 \rangle / \langle (X - \langle X \rangle)^2 \rangle^2$,
with $X \!=\! \bar\psi \psi$ or the Polyakov loop. 
At a second order transition, in the thermodynamic limit $B_4$ takes the value 1.604 dictated by the 
$3d$ Ising universality class.  
The critical lines bound
the quark mass regions featuring a chiral or deconfinement phase transition, and are called chiral and deconfinement critical lines, respectively.
The former has been mapped out on $N_t=4$ lattices \cite{fp3} and puts the physical quark mass
configuration in the crossover region.
The chiral critical line recedes strongly with decreasing lattice 
spacing \cite{LAT07, fklat07}: for $N_f=3$, on the critical point $m_\pi(N_t=4)/m_\pi(N_t=6)\sim 1.8$.  
Thus, in the continuum the physical point is much deeper in the crossover region
than on coarse lattices, \fig\ref{schem} (left). 

When a chemical potential is switched on, the chiral critical line sweeps out a surface, as shown
in \fig\ref{schem}. According to standard expectations \cite{derivs},
for small $m_{u,d}$, the critical line should 
continuously shift with $\mu$ to larger quark masses until it passes through the physical point at $\mu_E$, corresponding to the endpoint of the QCD phase diagram. 
This is depicted in \fig\ref{schem} (middle), where the critical point is part of 
the chiral critical surface. However, it is also
possible for the chiral critical surface to bend towards smaller quark masses, cf.~\fig\ref{schem} (right),
in which case there would be no chiral critical point or phase transition 
at moderate densities. 

With the currently available methods, there are then two ways of searching for signals of criticality.
One is to fix the quark masses to given, preferably physical values and calculate the effect of
switching on $\mu$. The other is to tune the quark masses to find the base line of the critical surface
and then follow its behaviour as a function of $\mu$. 

\section{The pseudo-critical temperature}

The first step of a calculation consists in determining the phase boundary.
In the last few years such calculations have become realistic for physical QCD. Since in this case 
the zero density transition is a smooth crossover, the phase boundary is only pseudo-critical even in 
the thermodynamic limit, and hence observable-dependent. It has a Taylor expansion
\be
\frac{T_c(\mu)}{T_c(0)}=1-\kappa(N_f,m_f)\left(\frac{\mu}{T}\right)^2+\ldots
\ee
The following two calculations with differently improved staggered fermions 
use the chiral condensate as an observable.
In \cite{tc_bi} the curvature  
was calculated with improved staggered quarks on $N_t=4,8$ lattices with a 
physical strange quark mass. Light quarks were extrapolated to the chiral limit assuming 
$O(4)$-scaling, giving $\kappa=0.059(2)(4)$. In another calculation\cite{tc_w} quark masses were
fixed to their physical values, so that no extrapolations and scaling assumptions had to be made. Simulations on $N_t=6,8,10$ lattices
were continuum extrapolated to yield 
$\kappa=0.066(20$. The important physics observation from these consistent results is that the curvature is only about a third of that of the
experimentally measured freeze-out curve \cite{freeze}, as shown schematically in \fig\ref{tc} (left).
Thus there appears a gap between the freeze-out curve and the QCD phase boundary.

\section{Signals for criticality at fixed mass}

\begin{figure}[t]
\vspace*{-7mm}
\hspace*{2mm}
\includegraphics[width=0.3\textwidth]{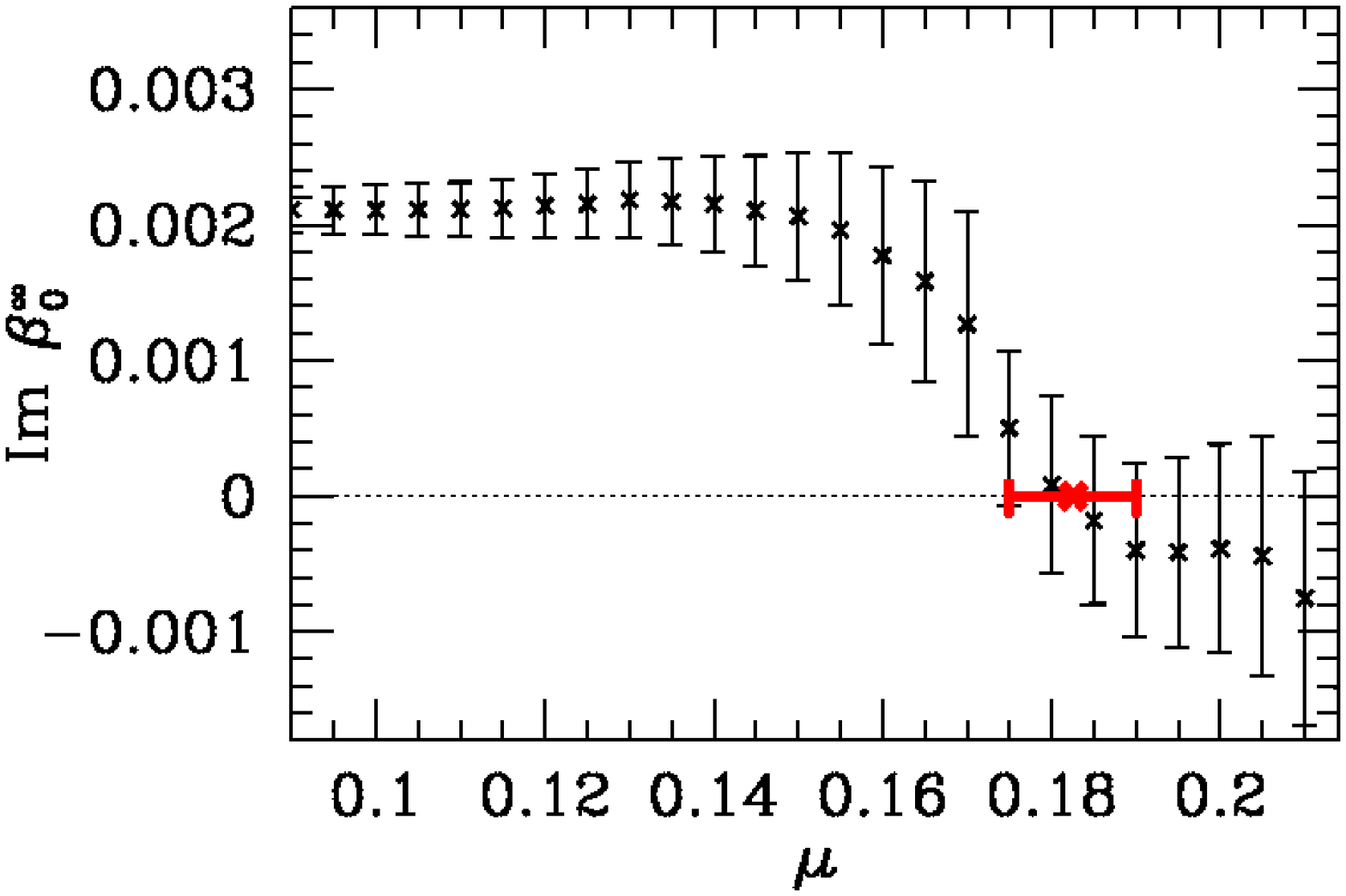}\hspace{3mm}
\includegraphics[width=0.3\textwidth]{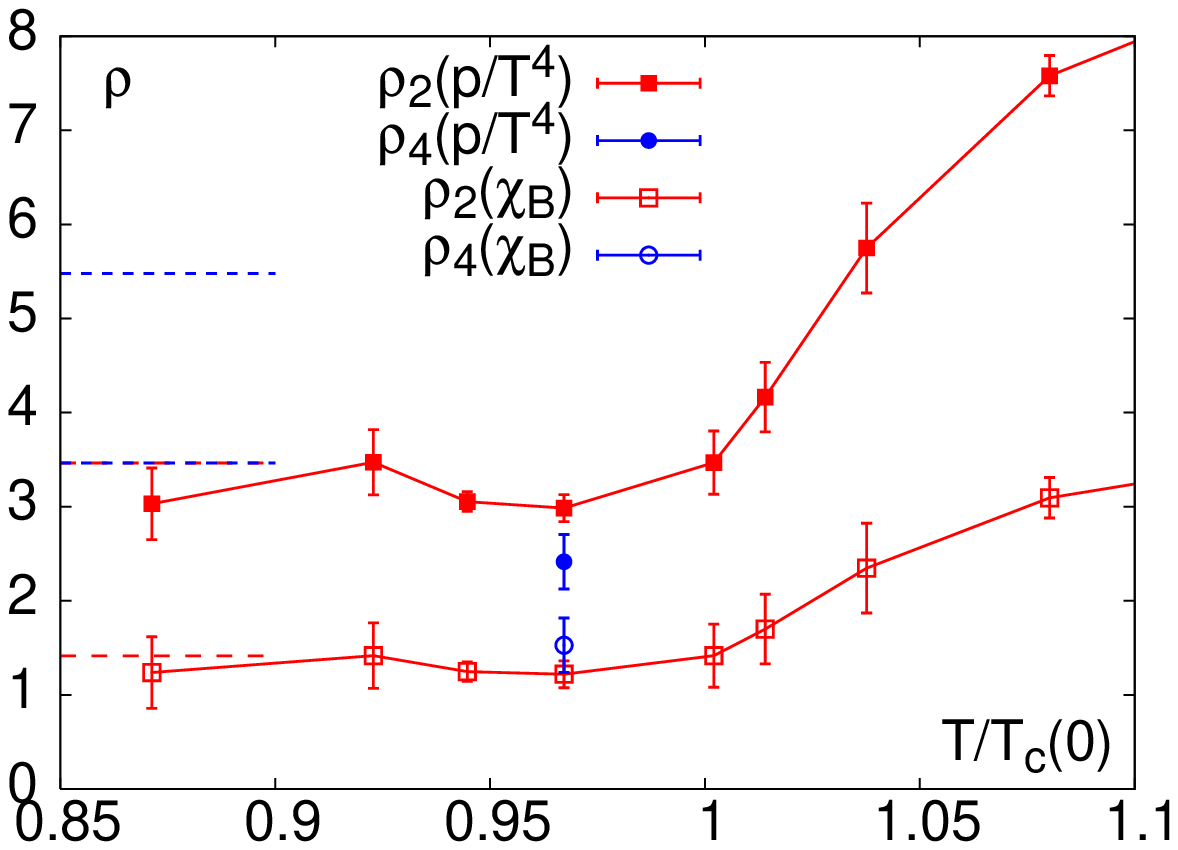}\hspace{3mm}
\includegraphics[height=0.2\textwidth]{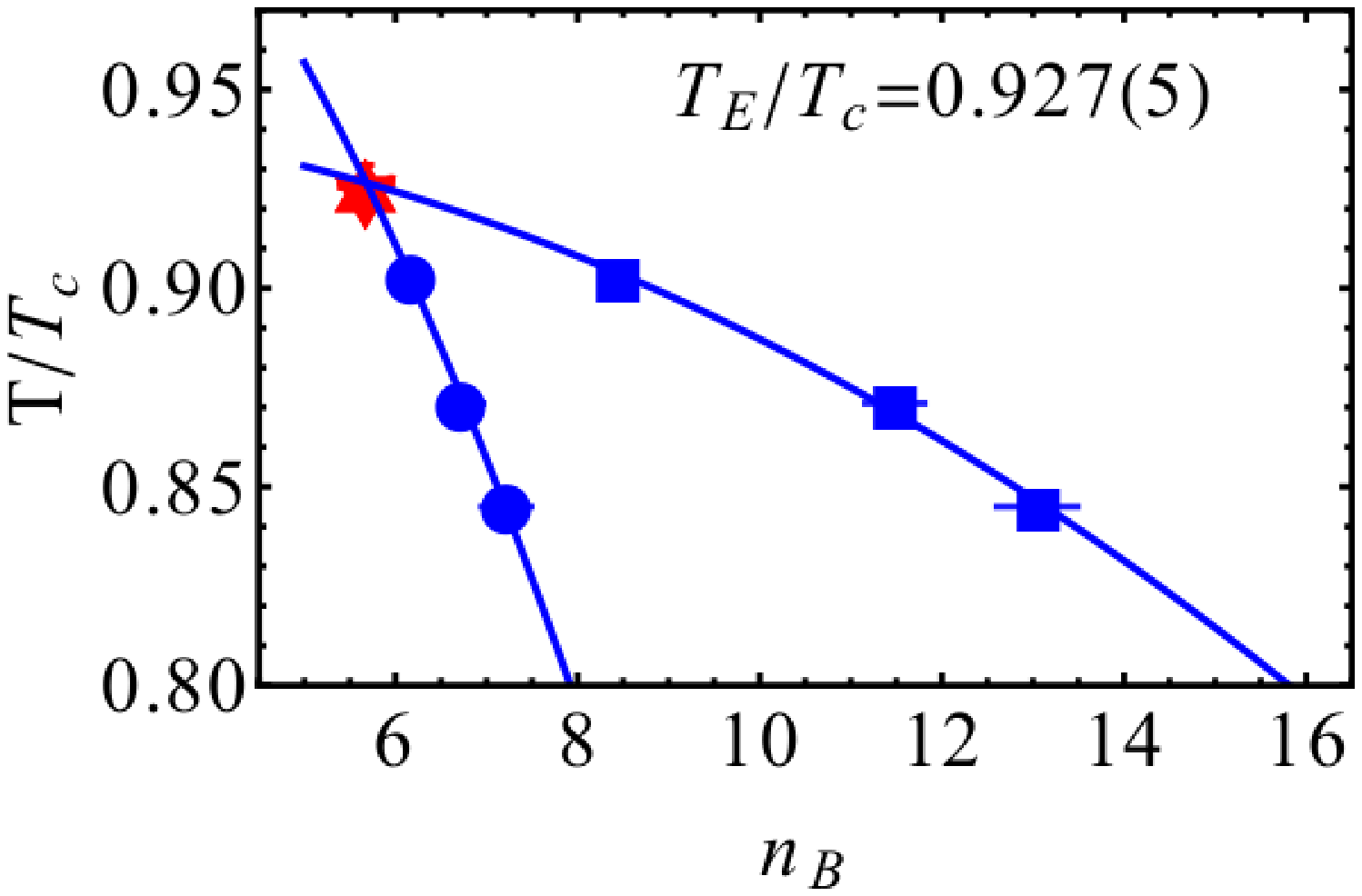}
\caption{\label{rew} Left: Imaginary part of the Lee-Yang zero closest to the real axis  \cite{fk}.
Middle: Estimates for the radius of convergence from different observables and at different
order for $N_f=2+1,N_t=4$. Dashed lines are hadron resonance gas values \cite{cs}. 
Right: First order coexistence region in baryon number density from the canonical ensemble on small $6^3\times 4$ lattices, $m_\pi\sim 700$ MeV \cite{liu}. }
\end{figure}

Reweighting methods at physical quark masses on $N_t=4$ lattices 
get a signal for a critical point 
at $\mu_B^E\sim 360$ MeV \cite{fk}. 
Quark masses were tuned 
to the mass ratios 
$m_{\pi}/m_{\rho}\approx 0.19, m_{\pi}/m_K\approx 0.27$, 
close to their physical values.
A Lee-Yang zero analysis was employed, as shown in \fig\ref{rew}. For a crossover the partition 
function $Z(V,\beta,N_t)$ has zeroes only for unphysical complex values of the lattice gauge coupling, whereas for a true phase transition the zeroes accumulate and pinch the real axis 
in the infinite volume limit.  
A caveat of this calculation is the observation that the critical point is found in the 
immediate neighbourhood of
the onset of pion condensation in the phase quenched theory, which is where
the sign problem becomes maximal \cite{kim}.
Therefore, it would be good to have a confirmation with an independent method.

One may also search for a critical point using the Taylor expansion.
In this case a true phase transitions will be signalled by an 
emerging non-analyticity, or a finite radius of convergence for the pressure series about $\mu=0$,
as the volume is increased, to be identified with the critical point,
\be
{p\over T^4}=
\sum_{n=0}^\infty c_{2n}(T) \left({\mu\over T}\right)^{2n}, \;\;
\frac{\mu_E}{T}=\lim_{n\rightarrow \infty}\rho_n, \;\;\rho_n=\sqrt{\left|\frac{c_{2n}}{c_{2n+2}}\right|},
\left|\frac{c_{0}}{c_{2n}}\right|^{1/2n}.
\label{press}
\ee
Theorems ensure that if the limit exists and asymptotically all coefficients of the series are positive, then there is a singularity on the real axis, which
would represent a critical point in the $(\mu,T)$-plane.
The current best attempt is based on four consecutive coefficients, 
i.e.~knowledge of the pressure to eighth order, and
a critical endpoint for the $N_f=2$ theory was reported  in \cite{Gavai}. 
There are also difficulties in this approach. Firstly, there are different definitions for the radius of convergence, which are only unique in the asymptotic limit, but differ quantitatively at finite order.
Furthermore, estimates at a given order are neither upper nor lower bounds on an
actual radius of convergence. Finally, finite estimates obtained at finite orders are a necessary, but not a sufficient condition for the existence of a critical point. For example, 
one also obtains finite estimates 
from the Taylor coefficients of the hadron resonance gas model, which does not 
feature a non-analytic phase transition. This is illustrated in \fig\ref{rew} (middle) \cite{cs}.

Another signal for a critical point \cite{liu} is based on simulations using the canonical ensemble of quark number $n$, which is the Fourier transform of the grand canonical partition function
evaluated at imaginary $\mu$, 
\be
Z_C(V,T,n)=\frac{1}{2\pi}\int d\phi\;e^{-in\phi}Z(V,T,i\mu)|_{\mu=i\phi}.
\ee
In this case the sign problem is deferred to the Fourier transform and restricts the total quark number   and hence the accessible volumes severely. The simulations in \cite{liu} are with Wilson fermions run on $6^3\times 4$ lattices with 
rather heavy pions, $m_\pi\sim700-800$ MeV. A first order coexistence region is detected
which merges at a critical point, whose baryon density can be converted to 
$\mu_B/T\sim 2.6$, \fig\ref{rew} (right). In this case there is no reweighting and no Taylor expansion 
involved. However, the lattice is coarse, its volume very small and the pion mass very far from physical, so again we cannot yet conclude for physical QCD. Because of the Fourier transform, the computational cost for extrapolations to the thermodynamic limit and to physical quark masses are growing nearly exponentially.

\section{The chiral critical surface}

\begin{figure}[t]
\vspace*{-0.5cm}
\includegraphics[width=0.37\textwidth]{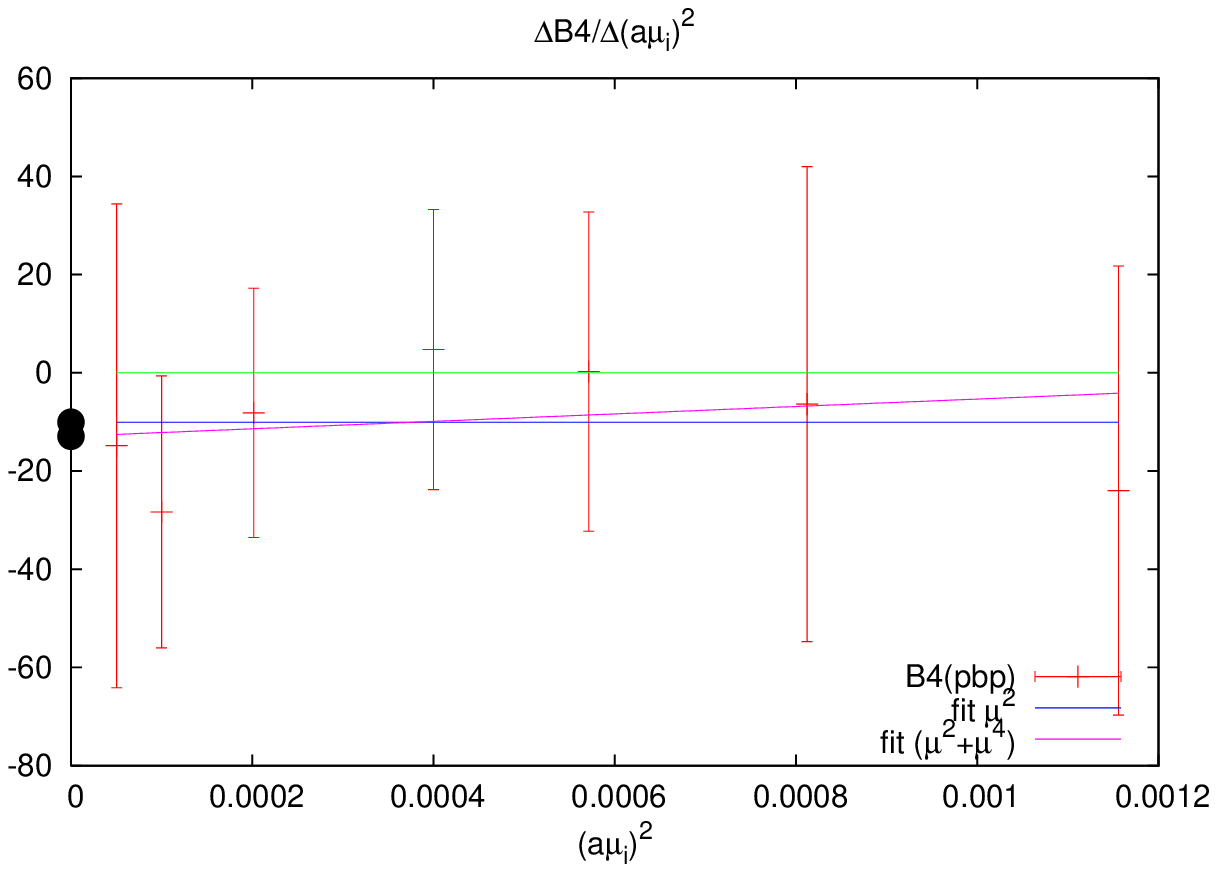}
\includegraphics[height=0.3\textwidth]{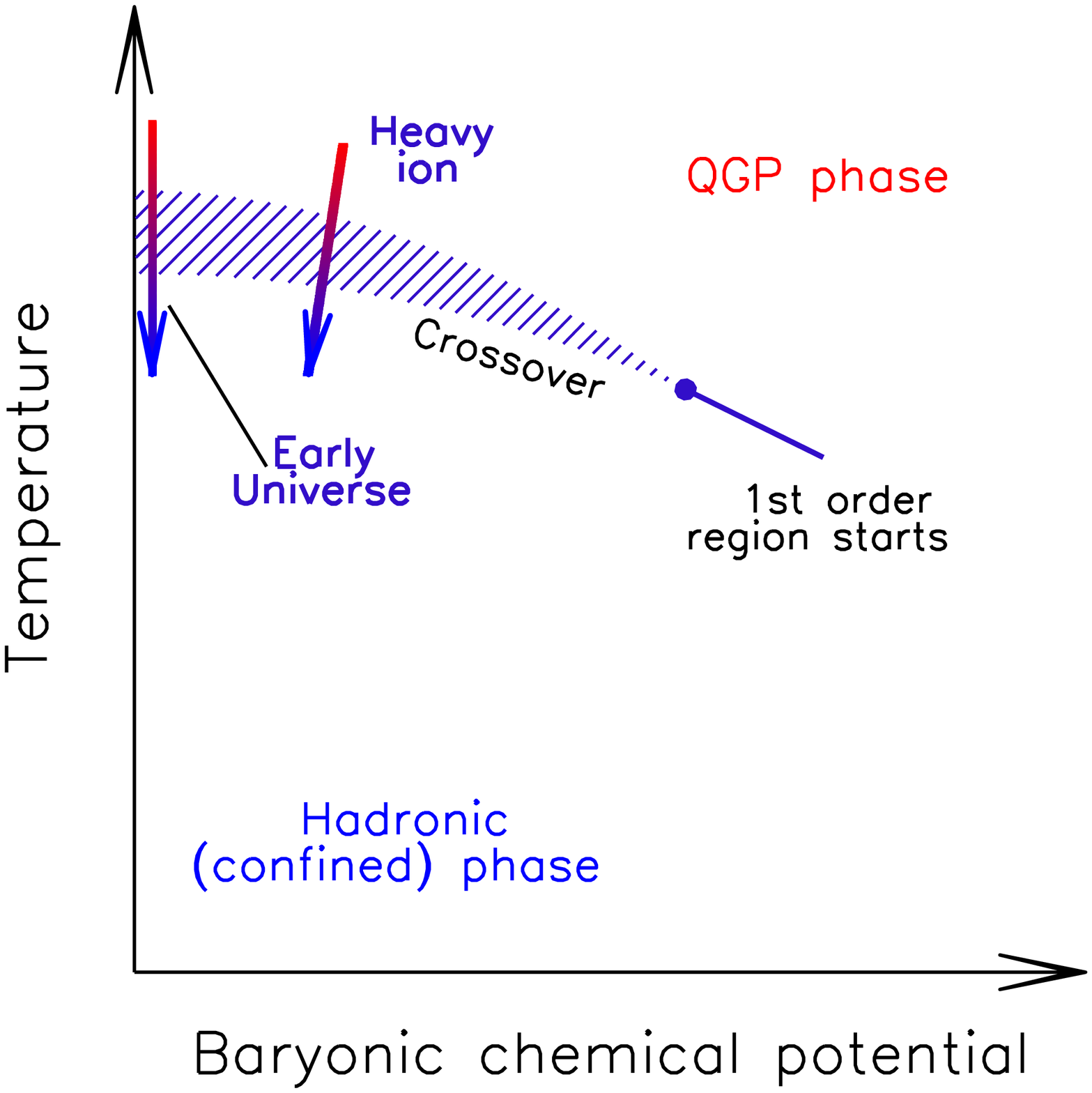}
\includegraphics[height=0.3\textwidth]{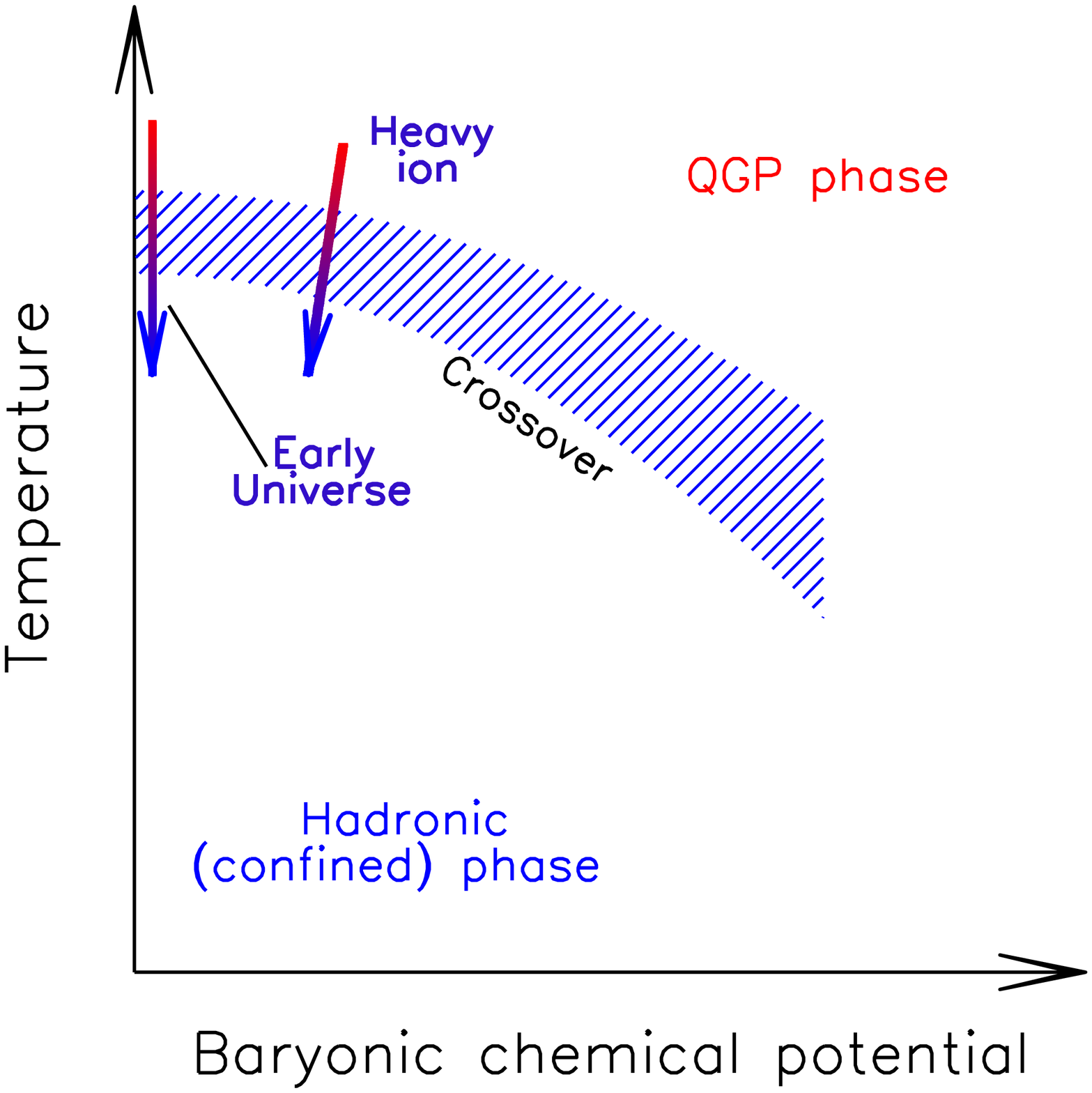}
\caption{\label{fkcurv} 
Leftt: Finite difference quotient for the curvature of the chiral critical surface, \eq(\ref{binder}), for $N_f=3$ on $N_t=6$. Middle=Right: The width of the crossover region should shrink to zero when approaching a critical point.
Instead it is found to widen with chemical potential, implying
a softening of the crossover \cite{tc_w}. }
\end{figure}

Rather than fixing to a definite set of quark masses, we shall now discuss the behaviour of the
chiral critical surface in \fig\ref{schem}.
For definiteness, let us consider three degenerate quarks, 
represented by the diagonal in the quark mass plane.
The critical quark mass corresponding to the boundary point of the chiral transition region
has an expansion
\be
\frac{m_c(\mu)}{m_c(0)}=1+\sum_{k=1}c_k \left(\frac{\mu}{\pi T}\right)^{2k}\,.
\ee
Tuning to $m_c(0)$, one may evaluate
the leading coefficients of this expansion. In particular, the sign of $c_1$ will tell us which of the scenarios
in \fig\ref{schem} is realised. 
The curvature of the critical surface in lattice units is directly related to the behaviour of the Binder cumulant via the chain rule,\be
\frac{dam_c}{d(a\mu)^2}=-\frac{\partial B_4}{\partial (a\mu)^2}
\left(\frac{\partial B_4}{\partial am}\right)^{-1}\,.
\label{binder}
\ee
In order to guard
against systematic errors, this derivative has been evaluated in two independent ways.
One is to fit the corresponding Taylor series of $B_4$ in powers of $\mu/T$ to data generated at 
imaginary chemical potential, the other to compute the derivative directly and without
fitting  \cite{fp3, fp4}. Both methods of calculation give mutually compatible results. 
After continuum conversion one finds for $N_f=3$ on $N_t=4$ that $c_1=-3.3(3), c_2=-47(20)$  \cite{fp4}.
The same behaviour is found for non-degenerate quark masses. Tuning the strange quark 
mass to its physical value,  we calculated
$m^{u,d}_c(\mu)$ with $c_1= -39(8)$ and $c_2<0$ \cite{grid}.
Hence, on coarse $N_t=4$ lattices, the region of chiral phase transitions shrinks, i.e.~the phase transition
weakens at least initially 
with a real chemical potential, and there is no chiral critical point for $\mu_B\lsim 500$ MeV, 
as in \fig\ref{schem} (right). This statement appears robust when a finer lattice is considered. 
As discussed in the zero density section, on $N_t=6$ the baseline
of the chiral critical surface moves closer to the zero mass origin, whereas its curvature remains
negative, \fig\ref{fkcurv} (left).

Indeed, the same observation can also be made at fixed masses. At zero density, the QCD transition
is just an analytic crossover, for which $T_c$ depends on the observable.
In particular, one may use this in order to define a width of the crossover region, for details see \cite{tc_w},
and study its behaviour as a function of chemical potential. If there is a critical point in the phase diagram,
the definition of $T_c$ becomes unique and the lines $T_c(\mu)$ computed from different observables
must meet at a critical point. In \cite{tc_w} the width of the crossover region was evaluated based on
the curvature of $T_c$, i.e.~its leading order Taylor coefficient, and extrapolated to the continuum, with a result shown schematically
in \fig\ref{fkcurv}. The width grows slightly rather than getting narrower, i.e.~the crossover gets
even softer initially as real chemical potential is switched on.
 
Note that one also observes a weakening of the phase transition with $\mu$ in the heavy quark case \cite{kim1} as well as a weakening of the transition with isospin chemical potential \cite{iso}, suggesting
that this is a generic feature for gauge theories with chemical potential for fermion number.

\section{Critical surfaces at imaginary chemical potential}

\begin{figure}[t]
\vspace*{-0.5cm}
\centerline{
\includegraphics[width=0.3\textwidth]{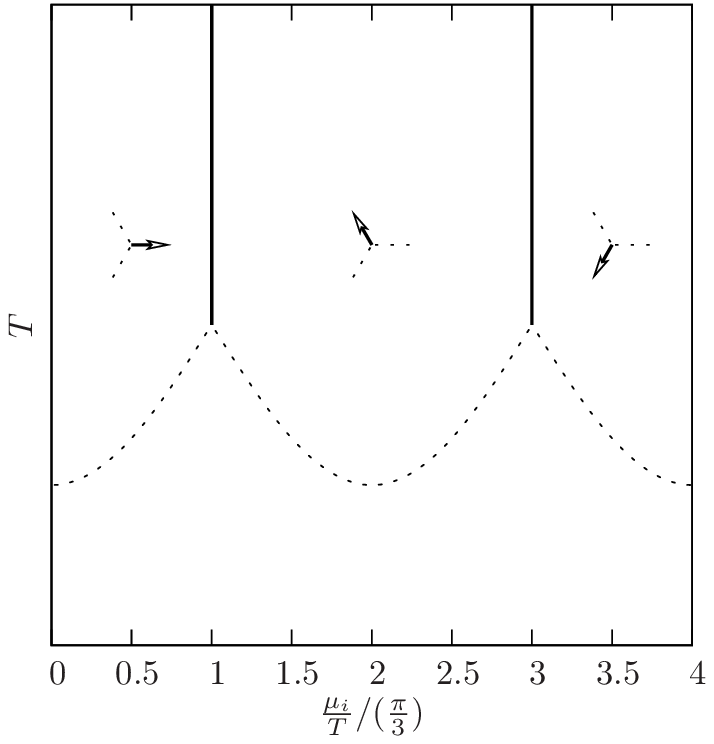}\hspace*{0.5cm}
\includegraphics[width=0.3\textwidth]{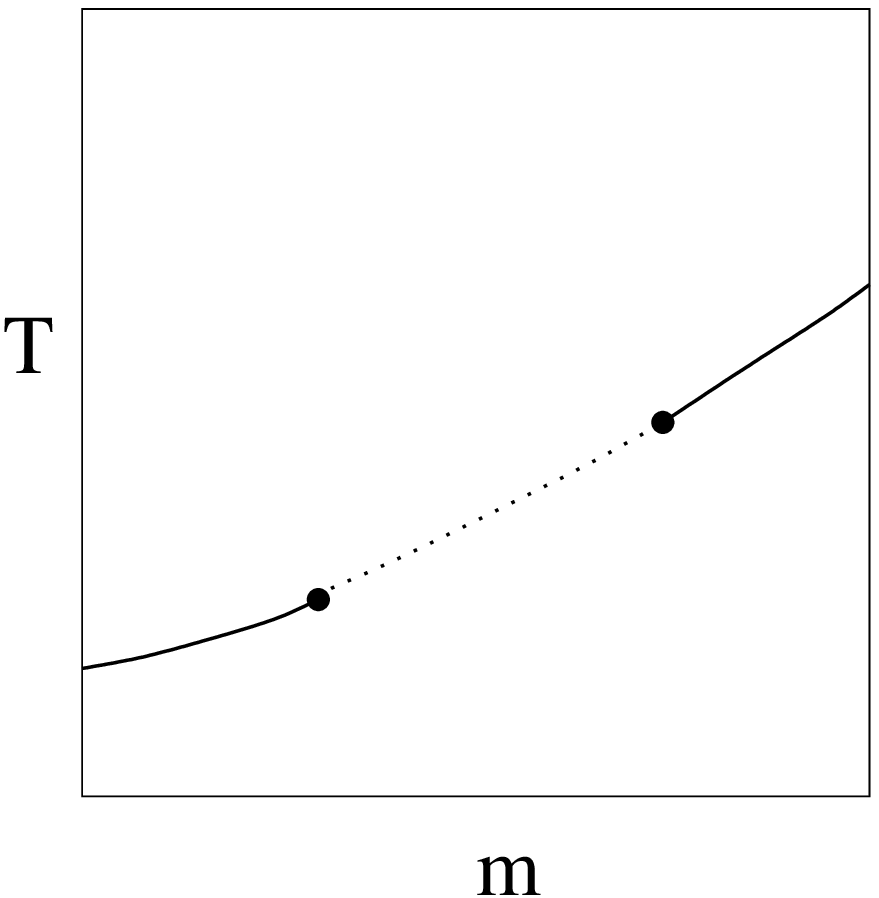}
\put(-85,80){$\langle \phi \rangle\neq 0$}
\put(-55,25){$\langle \phi \rangle =0$}\hspace*{0.5cm}
\includegraphics[height=0.3\textwidth]{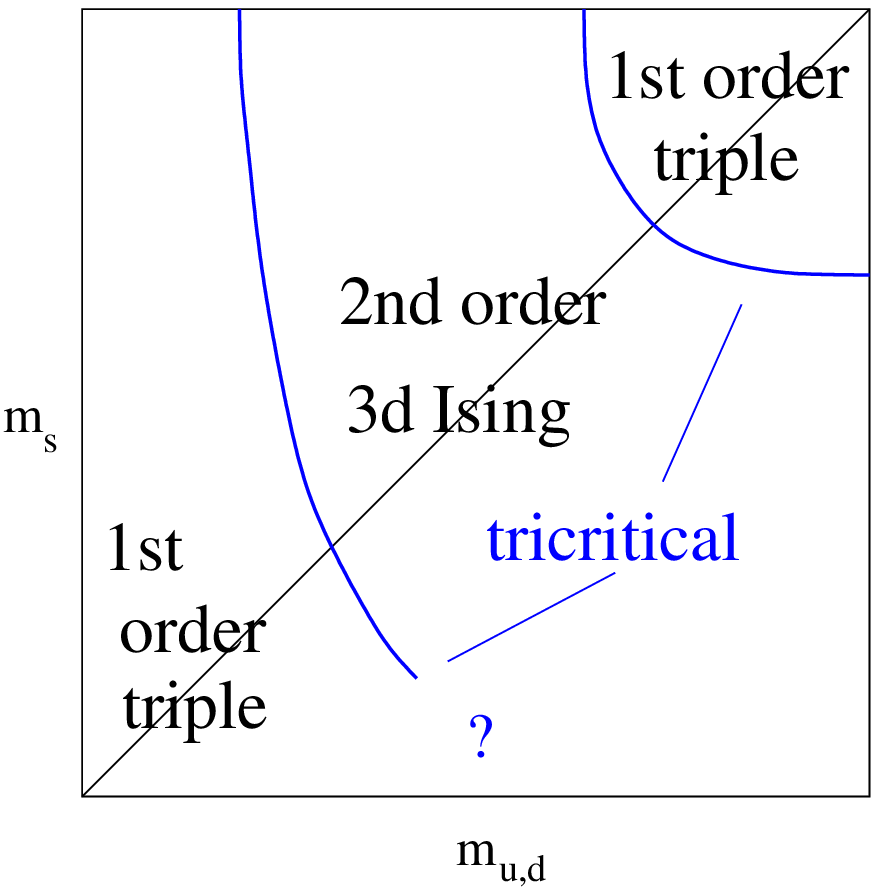}
}
\caption[]{Left: phase diagram for imaginary $\mu$. Vertical lines are
first order transitions between $Z(3)$-sectors, arrows show the phase
of the Polyakov loop. The $\mu\!=\!0$ chiral/deconfinement transition continues to imaginary $\mu$,
its order depends on $N_f$ and the quark masses.
Middle: Order of $Z(3)$ end point for $N_f=2,3$ at $\mu=i\pi T$. Solid lines are lines of triple points
ending in tricritical points,  connected by a $Z(2)$ critical line.
Right: Generalisation of the previous to non-degenerate quark masses.}
\label{schem1}
\end{figure}

Recent Monte Carlo studies at imaginary chemical potential indicate that at least for QCD 
this appears to be the general behaviour, and that the sign of the curvatures of the critical surfaces
can be understood. Moreover, since simulations at imaginary chemical potential have no sign problem,
they are completely controlled. As we shall see, there are interesting phase structures which can then 
be used to constrain model or effective theory descriptions of QCD.

The QCD partition function is cyclic in imaginary chemical potential, $Z(\mu/T) = Z(\mu/T+i2\pi n/3)$, due to its $Z(3)$ centre symmetry. This implies
transitions between adjacent centre sectors, distinguished by the phase of the Polyakov loop, 
at imaginary chemical potentials $\mu_i^c=(2n+1)\pi T/3$.
The schematic phase diagram is shown in \fig\ref{schem1}.
Transitions in $\mu_i$ between neighbouring sectors 
are of first order for high $T$ and analytic crossovers
for low $T$ \cite{rw,fp1}, as shown in \fig\ref{schem1} (left).
Correspondingly, for fixed $\mu_i=\mu_i^c$, there are transitions in $T$
between an ordered phase with two-state coexistence at high $T$ and a disordered phase at low $T$.
An order parameter to distinguish these phases is the
shifted phase of the Polyakov loop, $L=|L|\exp(i\varphi), \phi=\varphi-\mu_i/T$. 
At high temperature it fluctuates 
about $\langle \phi \rangle=\pm \pi/3$ 
on the respective sides of $\mu_i^c$. The thermodynamic limit picks one of those states,
thus spontaneously breaking the reflection symmetry 
about $\mu_i^c$. 
At low temperatures $\phi$ fluctuates smoothly between those values, with the 
symmetric ground state $\langle \phi\rangle =0$.
Away from $\mu_i=\mu_i^c$, there is a chiral or deconfinement
transition line separating high and low temperature regions. This line represents the analytic
continuation of the chiral or deconfinement  transition at real $\mu$. Its nature 
(1st, 2nd order or crossover) depends on the number of quark flavours and masses.  
The junction between the $Z(3)$ transition and the chiral/deconfinement transitions at fixed 
$\mu=i\pi T/3$ was studied in \cite{massimo, fprw} for $N_f=2,3$, respectively, with 
qualitatively similar results. For small/large quark masses the first order chiral/deconfinement transition
connects to the $Z(3)$ transition, and the junction is a triple point. For intermediate quark masses
they do not connect, and the $Z(3)$ transition has a $Z(2)$ end point. 
This results in the phase diagram for fixed $\mu_i^c$ shown in \fig\ref{schem1} (middle), which is 
qualitatively the same for $N_f=2,3$.

\begin{figure}[t]
\includegraphics[height=0.31\textwidth]{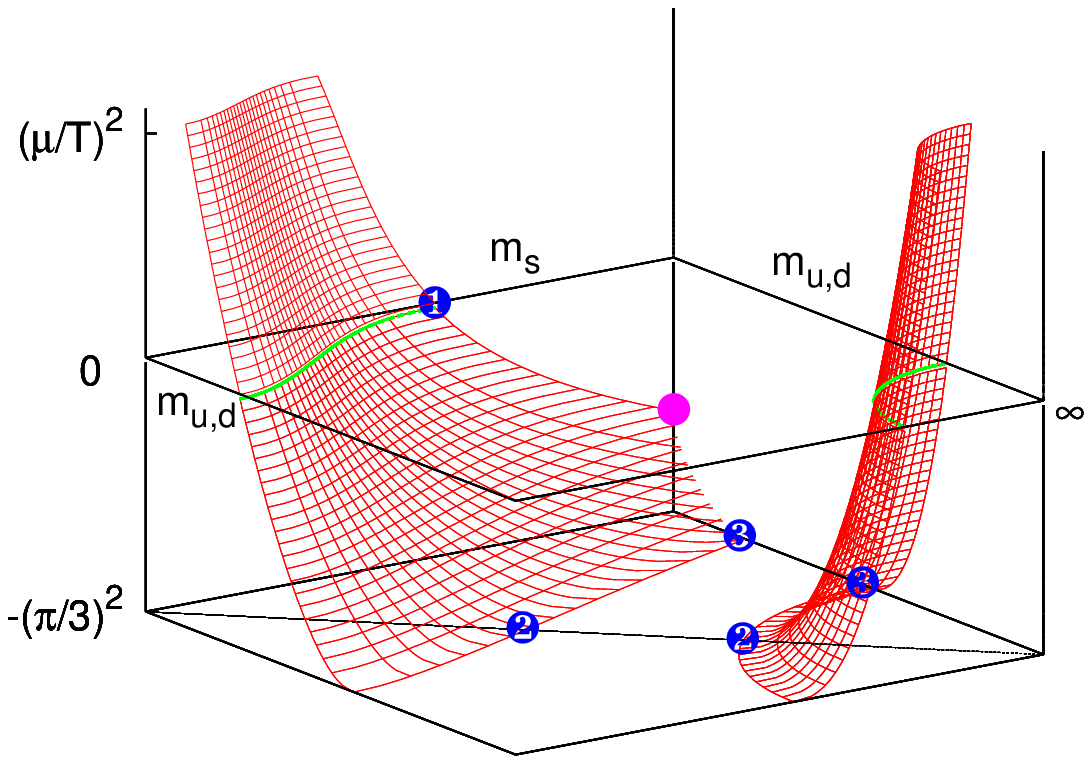}
\includegraphics[width=0.5\textwidth]{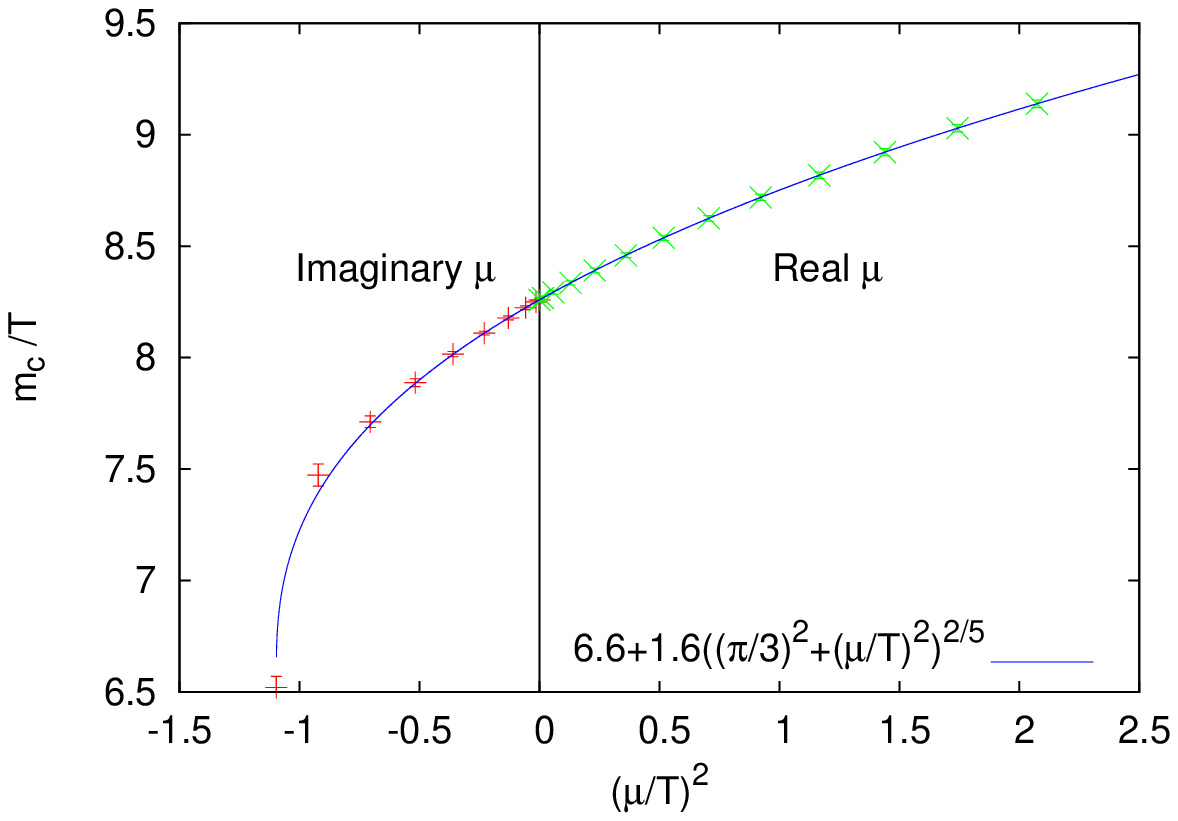}
\caption[]{\label{rw} 
Left: The critical surfaces delimiting the chiral and deconfinement transitions 
continue to imaginary, where they end in 
tricritical lines. Right: Critical line $m_c(\mu^2)$ in the 3-state Potts model fitted to \eq(\ref{mean})~\cite{kim1,fprw}.}
\end{figure}

The simplest generalisation to non-degenerate quark masses is then shown in \fig\ref{schem1} (right), where the respective tricritical points for $N_f=2,3$ are smoothly connected by tricritical lines, both in the
heavy and light mass regimes. Note that this diagram is the analogue for $\mu=i\pi T/3$ of the
diagram \fig\ref{schem} (left) for $\mu=0$. The key observation is now that 
the chiral and deconfinement critical surfaces continue to imaginary $\mu$ and 
terminate in the tricritical lines \cite{fprw}, as shown in \fig\ref{rw} (left).
In the case of heavy quarks, the critical surface is known over
a large range of imaginary and real $\mu$ within an effective theory \cite{kim1}, 
the 3d $Z(3)$ Potts model, which is
in the same universality class as QCD with heavy quarks. 
For fixed flavour content, i.e.~a slice through the critical surface, the deconfinement 
critical quark mass follows tricritical scaling \cite{fprw}, Fig.~\ref{rw} (right),
\be
\frac{m_c}{T}(\mu^2)=\frac{m_{\rm tric}}{T} + K \left[\left(\frac{\pi}{3}\right)^2+\left(\frac{\mu}{T}\right)^2\right]^{2/5}\;.
\label{mean}
\ee
The shape of the deconfinement critical surface is thus determined by the tricritical scaling law,  
while the sign of its curvature at $\mu=0$ follows from the fact that 
$m_\mathrm{tric}(\mu=i\pi T/3)<m_c(\mu=0)$.
While the chiral critical surface is not yet mapped out, one finds in the light quark mass regime $m_\mathrm{tric}(\mu=i\pi T/3)>m_c(\mu=0)$, which 
thus favours a negative sign for its curvature.
These findings are consistent with a monotonous weakening of the chiral and deconfinement transitions as $\mu^2$ gets more positive, which we now understand as being induced by the critical structure
at imaginary chemical potential.

\section{Conclusions}

In summary, signals from reweighting, radius of convergence estimates and canonical methods
are consistent with a critical point, but their systematics does not yet permit definite conclusions 
for physical QCD in the continuum.
On the other hand, following the chiral critical surface with comparatively controlled systematics tells us
that the chiral phase transition weakens with moderate $\mu$, thus leading us away
from the physical point. Possible scenarios are: if we mistrust the systematics of the former methods, 
we would conclude for either no critical point at all, or a critical point at larger chemical
potential $\mu_B\gsim 500$ MeV, where current methods break down.
However, another possibility is that all calculations hold, with a critical point at moderate densities
that would not belong to the chiral phase transition, but to physics unrelated to chiral symmetry breaking.
It would be interesting to distinguish these possibilities with the help of effective theories, which
can be gauged against the critical structures at imaginary chemical potentials.
\\  

\noindent
{\bf Acknowledgement:} This work is supported by the German BMBF, grant 06MS9150, and by the Helmholtz International Center for FAIR within the framework of the
LOEWE program by the State of Hesse.

\end{document}